\documentclass[12pt,a4paper,aps,preprint,superscriptaddress,nofootinbib]{revtex4-1}
\usepackage[utf8]{inputenc}
\usepackage{graphicx}
\usepackage{amssymb}
\usepackage{textcomp}
\usepackage{amsmath}
\usepackage{tabularx}
\usepackage{bm}
\usepackage{times}
\usepackage{color}
\usepackage{ulem}
\usepackage{slashed}
\usepackage{multirow}
\usepackage{verbatim}
\usepackage{cancel}
\usepackage{subfigure}
\usepackage{epstopdf}
\usepackage{mathrsfs}
\usepackage{amsmath}

\usepackage{tikz}
\usetikzlibrary{shapes,snakes}

\usepackage[export]{adjustbox}
\def\ba{\begin{align}}
\def\ea{\end{align}}
\def \eea {\end{eqnarray}}

\def \bea {\begin{eqnarray}}

\usepackage[colorlinks=true, pdfstartview=FitV, linkcolor=blue, citecolor=blue, urlcolor=blue]{hyperref}
\allowdisplaybreaks[4]

\def\lsim{\mathrel{\raise.3ex\hbox{$<$\kern-.75em\lower1ex\hbox{$\sim$}}}}
\def\gsim{\mathrel{\raise.3ex\hbox{$>$\kern-.75em\lower1ex\hbox{$\sim$}}}}

\definecolor{orange}{rgb}{1,0.5,0}

\begin{document}

%\title{$N_{\rm eff}$  and  the Pseudo-Dirac DM in Higgs portal}
\title{ Axion and Dark Fermion Electromagnetic Form Factors in Superfluid He-4}

\author{Wei Chao}
\email{chaowei@bnu.edu.cn}
\affiliation{
Center for Advanced Quantum Studies, Department of Physics,
Beijing Normal University, Beijing 100875, China
}

\author{Sichun Sun}
\email{sichunssun@bit.edu.cn}
\affiliation{School of Physics, Beijing Institute of Technology, Beijing, 100081, China}

\author{Xin Wang}
\email{xinwang17@mail.bnu.edu.cn}
\affiliation{
Center for Advanced Quantum Studies, Department of Physics,
Beijing Normal University, Beijing 100875, China
}
\author{Chen-Hui Xie}
\email{chenhuixie@mail.bnu.edu.cn}
\affiliation{
Center for Advanced Quantum Studies, Department of Physics,
Beijing Normal University, Beijing 100875, China
}

\begin{abstract}
Condensed matter materials have shown great potential in searching for light dark matter (DM) via detecting the phonon or magnon signals induced by the scattering of DMs off the materials.
In this paper, we study the possibility of detecting electromagnetic form factors of fermionic DM and axion-like particles (ALPs) using superfluid Helium-4.
The phonon induced by a sub-GeV fermionic DM scattering off the superfluid can be described using the effective field theory with the interaction between DM and the bulk ${}^4$He.  
Signals arising from the electromagnetic form factors of light DM in the presence of an external electric field are calculated.
Projected constraints on the charge radius, the anapole moment, and the magnetic moment of the DM are derived with 1 kg$\cdot$year exposure.  
The phonon signal induced by the scattering of ALPs off the superfluid is also calculated, which can put competitive and the first direct detection bounds on ALP-photon-dark photon couplings in the projected experiments.
\end{abstract}

\maketitle

%%%%%%%%%%%%%%%%%%%%%%%%%%%%%%%%
\section{Introduction}
\label{sec:Intro}
%%%%%%%%%%%%%%%%%%%%%%%%%%%%%%%%
Astrophysical observations have confirmed the existence of the neutral, non-baryonic, and weakly interacting cold dark matter (DM)~\cite{Bertone:2004pz},  which accounts for about 26.8\%~\cite{ParticleDataGroup:2020ssz} of the cosmic energy budget. Unfortunately, the Standard Model (SM) of particle physics,  which is consistent with almost all experimental observations, contains no DM candidate, except active neutrinos, which may serve as the hot DM candidate and cannot address the whole DM. These facts catalyze the birth of various DM models with DM mass ranging from $10^{-26}$ eV to the mass of celestial bodies.  Questions are how to directly detect the signal of DM in terrestrial laboratories and how to confirm the nature of DM. 

Traditional DM direct detection experiments~\cite{PandaX-4T:2021bab,XENON:2020kmp,CDEX:2018lau} aim to detect the recoil energy of nuclei or electrons arising from the collision of weakly interaction massive particle (WIMP) DM on the target in underground laboratories. Thanks to technological advancements, the direct detection sensitivity and efficiency have been greatly improved, and the exclusion limit on the WIMP-target nuclei scattering cross section almost reached the so-called neutrino floor~\cite{Monroe:2007xp,Strigari:2009bq,Billard:2013qya,Chao:2019pyh,OHare:2021utq},  the background from the coherent neutrino-nucleus elastic scattering. However, a confirmative DM signal is still not observed.  In recent years,  more and more attention are paid to the development of novel direct detection strategies using various condensed matter materials~\cite{Schutz:2016tid,Knapen:2016cue,Knapen:2017ekk,Griffin:2018bjn,Trickle:2019nya,Barbieri:1985cp,Trickle:2019ovy,Chigusa:2020gfs,Mitridate:2020kly,Trickle:2020oki,Mitridate:2023izi,Kahn:2021ttr}.   
In those proposed experiments,  signals of DM-nuclear scattering and DM-electron scattering can be regarded as nucleus and electron quasiparticles (phonon or plasmon), respectively.  
They are beneficial from both the low energy threshold and the high density of the solid-state system,  compared to atomic targets.  
Sub-GeV DM, which carries small kinetic energy and couldn't be detected in traditional direct detection experiments,  may be detected in these experiments considering that the excitation energy of a phonon is ${\cal O} (1)$ meV. 
We refer to Ref.~\cite{Kahn:2021ttr} for a conceptual review of DM direct detections in condensed matter. 

It has been demonstrated that a superfluid helium experiment would complement recently proposed superconductor experiments in detecting low-mass DM scattering on nucleons, with ${\cal O}(1)$ kg$\cdot$yr exposure~\cite{Knapen:2016cue,Guo:2013dt,Hertel:2018aal,Caputo:2019xum,Matchev:2021fuw,You:2022pyn,vonKrosigk:2022vnf,Baker:2023kwz,Anthony-Petersen:2023ykl,Hirschel:2023sbx,QUEST-DMC:2023nug}. 
The concept of using superfluid ${}^4$He as a detector was proposed and further developed in Refs.~\cite{Guo:2013dt,Hertel:2018aal,Knapen:2016cue,Maris:2017xvi,Knapen:2017ekk,Griffin:2018bjn,Hertel:2018aal,Coskuner:2018are,Acanfora:2019con,Lin:2019uvt,Caputo:2019cyg,Trickle:2019nya,Campbell-Deem:2019hdx,Caputo:2019ywq,Baym:2020uos,Kahn:2021ttr,Matchev:2021fuw,You:2022pyn,Murgui:2022zvy,Raya-Moreno:2023fiw,Caputo:2019xum}. 
Then, an effective field theory for superfluid ${}^4$He  has been developed to model the interactions among helium and quasiparticles, which allows a probe of sub-GeV DM.  
In this paper, we study the constraints of the proposed superfluid ${}^4$He  experiment on the electromagnetic form factors of Fermionic sub-GeV DM.    
With the help of the effective field theory (EFT) of interactions between DM and the bulk of ${}^4$He~\cite{Caputo:2019xum}, we calculate the cross-section of a sub-GeV Fermionic DM scattering off the Helium-4, emitting a single phonon and discuss the constraint of projected superfluid $^4$He experiment on the charge radius, the anapole moment and the magnetic moment of sub-GeV DM, which may arise from radiative corrections to the flavored DM~\cite{Bai:2013iqa,Bai:2014osa,Chao:2016lqd,Chao:2017emq}.  
We present the first constraint on these couplings in the condensed matter system and this study can be easily extended to the case of other condensed matter materials as the detector.

Considering that axion-like particles (ALPs) are promising DM candidates since they may address the strong CP problem and the DM problem simultaneously~\cite{Dine:1981rt,Preskill:1982cy,Abbott:1982af}, we study the phonon signal of the ALP in projected superfluid $^4$He experiment.  
Constraints on the ALP-photon-photon coupling ($g_{a\gamma\gamma}$) are usually given by Helioscope or Haloscope experiments~\cite{Irastorza:2011gs,McAllister:2017lkb}, in which an ALP is converted into a photon in the magnetic fields.  
It has been shown that there are exotic ALP models in which an ALP couples to the photon and a dark photon with coupling $g_{a\gamma\gamma^\prime}$.  
However, $g_{a\gamma\gamma^\prime}$ cannot be detected in any Helioscope experiment.
In this paper, we calculate the cross-section of an ALP scattering off the superfluid $^4$He into a dark photon and a phonon in the environment of an external electric field.  
Then we apply this formula to put a constraint on the coupling $g_{a\gamma \gamma^\prime}$ in projected superfluid $^4$He experiment with ${\cal O} (1)$ kg$\cdot$year exposure which is competitive to the other indirect experiments. It gives the first direct experimental constraint on this coupling.

The remaining paper is organized as follows: In section II we present the electromagnetic form factors of DM and briefly review the EFT for superfluid-DM interaction. Section III is devoted to the calculation of single phonon emission induced by the electromagnetic form factors of Fermionic DM. In section IV, we study the phonon signal induced by the ALP. The last part is the concluding remarks.

%%%%%%%%%%%%%%%%%%%%%%%%%%%%%%%%%%%%%%%%%%%%%%
\section{EFT FOR AXION-SUPERFLUID INTERACTION}
\label{sec:EFT}
Superfluid is a state of matter that behaves like a fluid with zero viscosity.  It is characterized by a spontaneously breaking $U(1)$ internal symmetry for a conserved number of atoms.  The Goldstone boson associated with the symmetry breaking corresponds to an acoustic phonon, which is a collective model equalling a density perturbation at long wavelengths. Phonon can be thought of as a stable quasiparticle that is essential for DM detection. Long-lived Phonons produced from low energy DM-superfluid scattering can be measured in ${\cal O}(1)$ kg superfluid helium when they propagate to the surface of the liquid.  In this paper, we focus on the direct detection signal of a Fermionic DM, $\chi$ and an ALP $a$.  Following the Ref.~\cite{Caputo:2019xum}, the most general low-energy effective action for this system can be written as
\bea
S_{\rm eff}^{} &=& \int d^4 x \left[ \bar \chi (i \slashed{\partial} - m_\chi) \chi -\alpha \bar \chi \gamma^\mu \chi \partial^\nu F_{\mu\nu}^{} -\beta^{} \bar \chi \gamma^\mu \gamma^5 \chi \partial^\nu F_{\mu\nu } - {\lambda \over 2 } \bar \chi \sigma^{\mu\nu} \chi F_{\mu\nu}^{}   \right. \nonumber \\
&&-{1\over 4} g_{a\gamma \gamma}^{} a F_{\mu\nu} \widetilde F^{\mu\nu} - {1\over 2 } g_{a\gamma \gamma^\prime}^{} a F_{\mu\nu}^{} \widetilde{F}^{\prime \mu\nu} + h.c.
\nonumber \\
&&+ \left. {1\over 4} F_{\mu\nu}^{} F^{\mu\nu}_{} -{1\over 2} a(X) F_{\mu\nu}^{} F^{\mu\nu}_{} - {1\over 2} b(X) F^{\mu \rho}_{} F^\nu_\rho \partial_\mu \phi \partial_\nu \phi  \right], \label{master}
\eea
where $F_{\mu\nu}^{}$  is the field strength of the photon, $\alpha$ is the charge radius, $\beta$ is the axial charge radius or anapole moment and $\lambda$ is the magnetic moment, $F^{\prime \mu\nu}$ is the field strength of the dark photon, $\phi \equiv \mu t + c_s \sqrt{\mu /\bar n} \pi $ being a scalar field for the spontaneous breaking of the U(1) with phonon  $\pi $ corresponding to the fluctuation of the field around the equilibrium configuration, $X=\sqrt{\partial_\mu \phi \partial^\mu \phi}$. 
Terms in the first and second lines of Eq. (\ref{master}) describe interactions of fermionic DM and ALP respectively, while the last two terms in the third line describe the coupling between photon and phonon.  
In the following analysis, we assume either $\chi$ or $a$ being the DM candidate, and we will not consider the case in which both $x$ and $a$ are coexisting DM.

It should be mentioned the electromagnetic form factors of DM may arise from radiative corrections to renormalizable DM interactions.  The electric dipole moment of DM is absent by assuming CP conserves DM interactions. We refer the reader to Refs.~\cite{Bai:2013iqa,Bai:2014osa,Chao:2016lqd,Chao:2017emq} for the calculation of these form factors.  Interactions of the ALP are usually derived from the triangle anomalies of gauge interactions~\cite{Marsh:2015xka,DiLuzio:2020wdo} and ALP may also couple to fermions. We will only focus only ALP couplings given in Eq.(\ref{master}) in the following analysis.

For the system at equilibrium, the last two terms reduce to the action for an electromagnetic field in a medium and one can use electric and magnetic polarizabilities $\alpha_E$ and $\alpha_B$ to fix the function $a$ and $b$: $a\sim 0$ and $ b\sim \bar n (\mu) \alpha_E^{} /\mu^2 $,  where $\bar n$ is the equilibrium number density and $\mu$ is the relativistic chemical potential.  Expanding $b$ and $\partial_\mu \phi \partial^\mu \phi$ to the next to leading order, one gets interactions between the phonon and the photon 
\bea
S_{int}^{} \sim \int d^4 x \sqrt{\mu \over \bar n } c_s  \left[ \left(  {\mu^2\over 2} {d b \over d\mu } +\mu b  \right)  \dot \pi F^{0\rho}_{} F^{0}_{~ \rho}  +\mu b \partial_j^{} \pi  F^{ij}_{} F_{0i}^{}  + \sqrt{\mu \over \bar n} c_s \partial_\mu^{} \pi \partial_\nu^{} \pi F^{\mu \rho}_{} F^{\nu}_{~\rho}  \right], \label{master2}
\eea
with which one may derive Feynman's rules for photon-phonon interactions. Here $c_s$ is the sound speed.  Relevant parameters for superfluid ${}^4$He are $\mu \approx m_{\rm He}^{} $, $c_s=248$ m/s, $\bar n=8.5\times 10^{22}~{\rm cm}^{-3}$~\cite{abraham},  $ \alpha_E \approx 2\times 10^{-25} {\rm cm}^3$~\cite{Fetter} and $ \mu^2 db /d\mu \approx {\alpha_E \bar n / m_{\rm He } c_s^2 }$~\cite{Caputo:2019xum}, according to the thermal dynamical identities.  
Eq.~(\ref{master2}) shows that there are two photons in each interaction. Thus introducing an external electron field allows the conversion of a photon into a phonon, which is analogous to the Primakoff effect~\cite{Raffelt:1985nk,Raffelt:1987np}. The Feynman rule for photon-phonon transition in an external electric field is given in Ref.~\cite{Caputo:2019xum}, which will be applied to the calculation here.

\section{Phonon induced by the Electromagnetic form factors of dark fermion}

In this section, we focus on the analytical calculation of the dark fermion superfluid ${}^4$He scattering. 
According to the action in Eq.~(\ref{master2}), the Feynman rule for the photon-phonon mixing in the presence of an external electric field $E_i$ can be written as
\bea
<\pi|\mathcal{L}_{eff}|A_\mu>=\frac{\partial\mathcal{L}_{eff}}{\partial\pi\partial\,A_\mu}=\frac{1}{2}\frac{\alpha_E\bar{n}}{m_{He}c_s}\sqrt{\frac{m_{He}}{\bar{n}}}E^i\omega_q(\omega_q\delta^\mu_i+q_i\delta^\mu_0), \label{master3}
\eea 
where $\omega_q$ and $q_i$  are the energy and momentum of the photon, respectively.  
With Eq.~(\ref{master3}) we may calculate the phonon signal induced by electromagnetic magnetic form factors of Fermionic DM that passes through a superfluid in the environment of external electric fields.  
In the following, we will study the contributions of various electromagnetic form factors of $\chi$ to the phonon signal, separately.  
Results are given in both the relativistic limit and the non-relativistic limit.

\subsection{ Non-relativistic limit}
\begin{figure}
	\begin{center}
		\begin{tikzpicture}[scale=2]
			\draw[-,ultra thick] (0,2)--(1,1);
			\draw[-,ultra thick] (0,0)--(1,1);
			\draw[-stealth,ultra thick] (0,2)--(0.5,1.5) ;
			\draw[-stealth,ultra thick] (1,1)--(0.5,0.5) ;
			\draw[-,snake=snake, ultra thick] (1,1) -- (1.9,1);
			\draw[-,ultra thick] (2,1) circle (0.1) ;
			\draw[-,  thick, rotate around={45:(2,1)}] (1.9,1) -- (2.1,1);
			\draw[-,  thick, rotate around={-45:(2,1)}] (1.9,1) -- (2.1,1);
			\draw[dashed, ultra thick] (2.1,1) -- (3,1);
			\node[red, thick] at (0.5,2.3) {$k$};
			\node[red, thick] at (0.5,-0.3) {$k'$};
			\node[red, thick] at (2.5,0.6) {$q$};
		\end{tikzpicture}
		\caption{Feynman diagram of DM and superfluid helium scattering producing a phonon\label{fig:feynpro-rxx}}
	\end{center}
\end{figure}
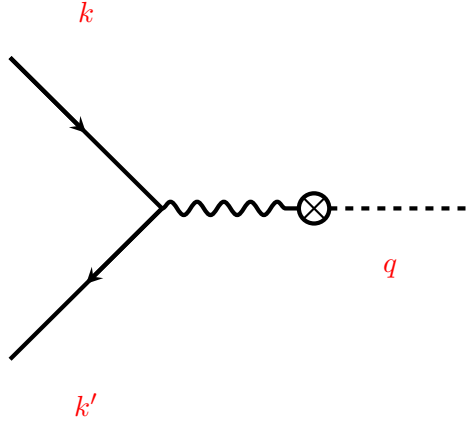

According to the action in Eq. (\ref{master}), we can derive the Feynman rule for the vertex of  the charge radius of $\chi$ as
\bea
\langle A_\rho,\bar{\chi}|\mathcal{L}|\chi\rangle=\frac{\partial\mathcal{L}}{\partial\chi\partial\bar{\chi}\partial\,A_\rho}=\alpha^{} \gamma^\mu(q_\mu\,q_\rho-q^2g_{\mu\rho})~.
\eea 
When $\chi$ passes through the superfluid in the presence of an external electric field, the scattering will excite the $^4$He producing a phonon. which can be described by the Feynman diagram given in Fig.~\ref{fig:feynpro-rxx}.
The scattering amplitude induced by the charge radius is 
\bea
i\mathcal{M}&=&\bar{u}(-i\alpha^{}\gamma^\mu)u(q_\mu\,q_\nu-q^2g_{\mu\nu})\frac{-ig^{\nu\rho}}{q^2}i\frac{\bar{n}\alpha_E}{m_{He}\,c_s}\sqrt{\frac{m_{He}}{\bar{n}}}\omega_qT_\rho\nonumber
\\&=&\bar{u}i\alpha\gamma^\mu\,u\frac{\bar{n}\alpha_E}{m_{He}\,c_s}\sqrt{\frac{m_{He}}{\bar{n}}}\omega_qT_\mu~,\label{eq:M1}
\eea
where $\bar u(k^\prime )$ and $u(k^{})$ are spinors of DM, $q$ is the momentum of the outgoing phonon, $T_\mu$ is a four-vector defined as 
\bea
T=(\textbf{E}  \cdot \textbf{q} , q^0 \textbf{E})~,
\eea
with $\textbf{E}$ the field strength of the external electric field, which has $T\cdot q=0$.   In the non-relativistic limit, where
 $dcos\theta=\frac{d\omega}{v_\chi|\mathbf{q}|}$, also we have these relations $ k' = k = (m_\chi,\textbf{0} )$,  $q^2 =-|\textbf{q}|^2$. The differential scattering rate can then be written as
\bea
{{d\Gamma } \over {d\omega }} = {{\bar n\alpha _E^2{\alpha ^2}} \over {4\pi {\mkern 1mu} c_s^6{m_{He}}{v_\chi }}}|{\textbf{E}}{|^2}\omega _q^4co{s^2}{\theta _E}~,
\eea 
where {{$\cos\theta_E=\cos\theta\cos\theta_\chi-\cos(\phi-\phi_\chi)\sin\theta\sin\theta_\chi$\cite{Caputo:2019xum}}} with $\theta_E$ the angle between the external electric field and the outgoing phonon, {{and $(\theta_\chi,\phi_\chi)$ the angle between the incoming DM and the electric field}}.  {{Moreover, $\cos\theta=\frac{c_s}{v_\chi}+\frac{q}{2m_\chi v_\chi}$ is the angle between the incoming DM and the outgoing phonon}}. $v_\chi$ is the velocity of DM. Notice that the rate can be enhanced by a large field strength of the external electric field.  
\\The event rates per unit target mass are obtained by~\cite{DelNobile:2021wmp,Griffin:2019mvc}
\begin{align} \label{eq:R}
	R=\int dv_\chi\,f_\text{MB}(v_\chi)\frac{\rho_\chi}{m_\text{He}\bar n m_\chi}\int_{\omega_\text{min}}^{\omega_\text{max}}d\omega \frac{d\Gamma}{d\omega}\,.
\end{align}
where $\rho_\chi$ is the local energy density of the  DM in our solar system. Here we take $\rho_\chi=0.4\,~\text{GeV}/\text{cm}^3~$\cite{Trickle:2020oki}. The dark matter Maxwell-Boltzmann distribution in the Milky Way halo is given by~\cite{Acanfora:2019con}
\begin{align}
	f_{MB}(v_\chi)=4\frac{v_\chi^2}{v_0^2}\frac{e^{-v_\chi^2/v_0^2}\Theta(v_{esc}-v_\chi)}{\sqrt{\pi}v_0 erf\left(\frac{v_{esc}}{v_0}\right)-2e^{-v_{esc}^2/v_0^2}v_{esc}}~, \label{eq:MB}
\end{align}                                                                                               
with $v_0\simeq220$~km/s and $v_{esc}\simeq550$~km/s~ being the escape velocity of DM.
The event rate per unit target mass is then 
\bea
R_{\alpha}^{} &=&\int_{0}^{v_{esc}}\frac{1}{N_0}e^{-\frac{v_\chi^2}{v_0^2}}v_\chi^2dv_\chi4\pi\frac{\rho_\chi}{m_{He}\bar{n}m_\chi}\frac{\bar{n}\alpha_E^2\alpha^2}{4\pi\,c_s^6m_{He}v_\chi}|\textbf{E}|^2\int\omega^4cos^2\theta_Ed\omega \nonumber\\
&=&\frac{\rho_\chi\alpha_E^2\alpha^2|\textbf{E}|^2}{N_0m_{He}^2m_\chi\,c_s^6}\int^{v_{esc}}ve^{-\frac{v^2}{v_0^2}}dv\int\omega^4cos^2\theta_Ed\omega~, \label{master4}
\eea 
%Here  $v_{esc}^{}=550$~km/s~ . We have taken the DM velocity distribution as the Maxwell-Boltzmann distribution. 

According to the Lagrangian in the Eq.~(\ref{master}),  the Feynman rule for the vertex of the anapole moment is the same as that of the charge radius up to the replacement: $\gamma^\mu \to \gamma^\mu \gamma_5$.
%\begin{align}
%	\mathcal{L}&=\beta\bar{\psi}\gamma^\mu\gamma^5\psi\partial^\nu\,F_{\mu\nu}~.\label{eq:L2}
%\end{align}
Following the same procedure, the contribution of the anapole moment to the scattering amplitude
\bea
i\mathcal{M}&=&\bar{u}(-i\beta\gamma^\mu\gamma^5)u(q_\mu\,q_\nu-q^2g_{\mu\nu})\frac{-ig^{\nu\rho}}{q^2}i\frac{\bar{n}\alpha_E}{m_{He}\,c_s}\sqrt{\frac{m_{He}}{\bar{n}}}\omega_qT_\rho\nonumber
\\&=&\bar{u}i\beta\gamma^\mu\gamma^5u\frac{\bar{n}\alpha_E}{m_{He}\,c_s}\sqrt{\frac{m_{He}}{\bar{n}}}\omega_qT_\mu \; .\label{eq:M2}
\eea 
In the non-relativistic limit, one has $k^2 T^2\simeq \ m_\chi^2(\textbf{E}\cdot\textbf{q})^2=(k\cdot\ T)^2=(k'\cdot\ T)^2$, where we have assumed $\omega\ll|\textbf{q}|$. 
One can easily check that the squared matrix element for this process is zero in the non-relativistic limit. 

Now we turn to the calculation of magnetic moment induced phonon signal. The Feynman rule for the vertex of  the magnetic moment  is 
\bea
\langle \bar{\chi}A_\rho|\mathcal{L}|\chi \rangle =\lambda \sigma^{\mu\rho}q_\mu~,
\eea 
so the scattering amplitude is 
\bea
i{\cal M} =\bar{u}(-i\lambda\sigma^{\mu\nu}q_\mu)u\frac{1}{q^2}\frac{\bar{n}\alpha_E}{m_{He}\,c_s}\sqrt{\frac{m_{He}}{\bar{n}}}\omega_qT_\nu \; ,\label{eq:M3} 
\eea 
which results in the following event rate
\bea
\frac{d\Gamma}{d\omega}=\frac{\bar{n}\alpha_E^2\lambda^2}{4\pi\,c_s^4m_{He}v_\chi}\omega_q^4|\mathbf{E}|^2cos^2\theta_E  \; . 
\eea 
Thus the event rate induced by the magnetic moment of the DM per unit target mass  can be written as 
\bea
R_{\lambda}^{} =\frac{\rho_\chi\alpha_E^2\lambda^2|\textbf{E}|^2}{N_0m_{He}^2m_\chi\,c_s^4}\int^{v_{esc}}ve^{-\frac{v^2}{v_0^2}}dv\int\omega^4cos^2\theta_Ed\omega~. \label{master5}
\eea
Using Eqs.~(\ref{master4}) and (\ref{master5}) one may estimate of constraints of projected superfluid ${}^4$He experiment on the electro-magnetic form factors of light Fermionic DM, which will be the first constraint on these parameters with the condensed matter material detectors. Similarly one may also derive the constraint of the dipole moment of DM, which will be presented in an incoming work.

\begin{figure}[t]
	\centering
	\includegraphics[width=0.7\textwidth]{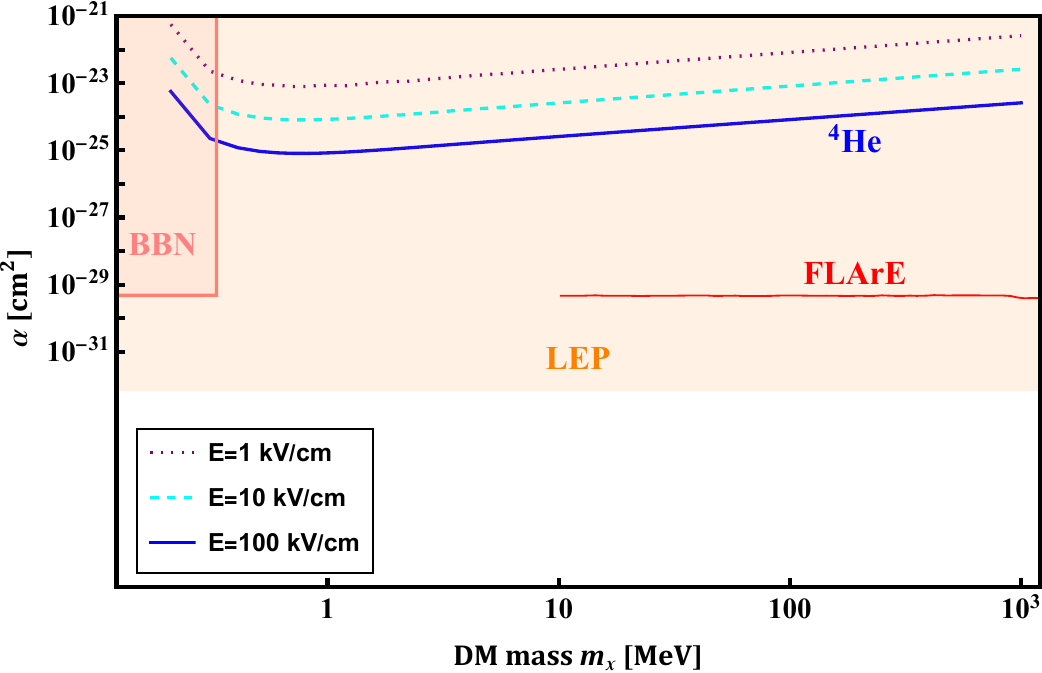}
	\caption{constraints on charge radius when $E$=1 , 10 , 100 ~kV/cm.  We consider the 95$\%$ C.L. exclusion limit for 1 kg$\cdot$year exposure  of the superfluid $^4$He, assuming zero background.  Shaded constraints are taken  from LEP~\cite{Chu:2018qrm}, FLArE~\cite{Kling:2022ykt} and BBN~\cite{Chu:2019rok}, respectively.} 
	\label{fig:alpha}
\end{figure}

\subsection{The general case}

In this subsection, we calculate the DM-Superfluid-${}^4$He scattering in the general case, where we take $k=(E,\textbf{k})$ as the four-momentum of incoming DM, $q=(\omega,\textbf{q})$ as the four-momentum of the outgoing phonon and $k'=(E',\textbf{k}')$ as the four-momentum of the outgoing DM. According to the  dispersion relation $E'^2=m_\chi^2+|\textbf{k}'|^2$,  the expression of cos$\theta$  can be given as
\begin{align}
	cos\theta=\frac{c_s}{v_\chi}+\frac{\omega}{2c_s|\textbf{k}|}-\frac{c_s\omega}{2|\textbf{k}|}~,
\end{align}
where $\theta$ represents the angle between the incoming DM and the phonon, $v_\chi$ is the velocity of the incoming DM, and $|\textbf{k}|$ is the module of momentum of the incoming DM. We  can also get the derivation of $\cos\theta$, $dcos \theta  = {{d\omega } \over {{v_\chi }|\textbf{q}|}}$, where $|\textbf{q}|$ is the module of the momentum of phonon. We have used the dispersion relation in superfluid ${}^4$He, $w=c_s|\textbf{q}|$ \cite{Matchev:2021fuw,Schmitt:2014eka}.

As mentioned before, the Lagrangian with charge radius and its amplitude correspond to Eq.~\eqref{eq:L1} and Eq.~\eqref{eq:M1}, respectively.
One gets the differential emission rate
\begin{align}
	\frac{d\Gamma}{d\omega}=\frac{\alpha^2\bar{n}\alpha_E^2\omega^4|\textbf{E}|^2}{8\pi\,m_\chi\,m_{He}c_s^6v_\chi\,E}\Bigg[2(Ecos\theta_E-|\textbf{k}|c_scos\theta_\chi)^2+(cos^2\theta_E-c_s^2)\Big(\omega\,E-\frac{\omega|\textbf{k}|}{c_s}cos\theta\Big)\Bigg]~.
\end{align}
{{The definitions of angles $(\theta_\chi,\phi_\chi)$ and $\cos\theta_E$ here are the same as those defined in the previous subsection}}.
%$(\theta_\chi,\phi_\chi)$ is the angle between the incoming DM and the electric field. The angle between the electric field and phonon is described by $\cos\theta_E=\cos\theta\cos\theta_\chi-\cos(\phi-\phi_\chi)\sin\theta\sin\theta_\chi$\cite{Caputo:2019xum}.
Similarly, one gets the differential emission rate with anapole moment
\begin{align}
	\frac{d\Gamma}{d\omega}=\frac{\beta^2\bar{n}\alpha_E^2\omega^4|\textbf{E}|^2}{8\pi\,m_\chi\,m_{He}c_s^6v_\chi\,E}\Bigg[2(Ecos\theta_E-|\textbf{k}|c_scos\theta_\chi)^2+(cos^2\theta_E-c_s^2)\Big(\omega\,E-\frac{\omega|\textbf{k}|}{c_s}cos\theta-2m_\chi^2\Big)\Bigg].
\end{align}
The differential emission rate for the magnetic moment becomes
\begin{align}
	\frac{d\Gamma}{d\omega}&=\frac{\lambda^2\bar{n}\alpha_E^2|\textbf{E}|^2\omega^2}{8\pi\,m_\chi\,m_{He}c_s^4v_\chi\,E}\Bigg[-2(Ecos\theta_E-|\textbf{k}|c_scos\theta_\chi)^2+\notag
	\\&(cos^2\theta_E-c_s^2)\Big(-2c_s^2E^2+4c_sE|\textbf{k}|cos\theta-2|\textbf{k}|^2cos^2\theta+2m_\chi^2+\omega\,E-\frac{\omega|\textbf{k}|}{c_s}cos\theta\Big)\Bigg]~.
\end{align}
Given the expression of differential rates,  one can get the event rates per unit target mass by numerical computation.

\subsection{Numerical results}

To carry out numerical analysis, we need first to confirm the kinematics of the scattering process. 
According to $\cos\theta\le1$, one gets $0<\omega\le\,2c_sE(v_\chi-c_s)$. 
Since the derived event rates are only effective for phonon, we have the maximum energy limit  $\omega_\text{max}=\min\big(2E c_s(v_\chi-c_s),c_s\Lambda\big)$, where $\Lambda$ is a UV cutoff of the superfluid
EFT set as $\Lambda\sim {\cal O}(1)$~keV.
For the detector threshold,  calorimetric techniques have a sensitivity at 1 meV~ \cite{Hochberg:2015pha} and the quantum evaporation of the helium atom~\cite{Hertel:2018aal,Maris:2017xvi} sets the minimum energy of phonon as  $\omega_\text{min}=0.62$~meV~\cite{Maris:2017xvi}, which is required by a phonon to overcome the binding energy between ${}^4$He atoms and the remaining bulk material.
Given these, we can compute the projected excluded limit on the electromagnetic form factors assuming a null phonon signal with certain exposure in a projected experiment. We show in Fig.~\ref{fig:alpha}  the exclusion limit on the charge radius of the DM at the 95\% C.L. with ${\cal O}(1)$ kg$\cdot$year exposure, where the solid, dashed, and dotted lines correspond to $E$=1, 10, 100 ~kV/cm, respectively. 
We can see that when the electric field strength increases by one magnitude, the limit of the charge radius is one order of magnitude stronger. Other shaded constraints are taken  from LEP~\cite{Chu:2018qrm}, FLArE~\cite{Kling:2022ykt} and BBN~\cite{Chu:2019rok}, respectively.

\begin{figure}[t]
	\begin{center}
		\begin{minipage}[c]{0.5\textwidth}
			\includegraphics[height=6cm,width=8cm]{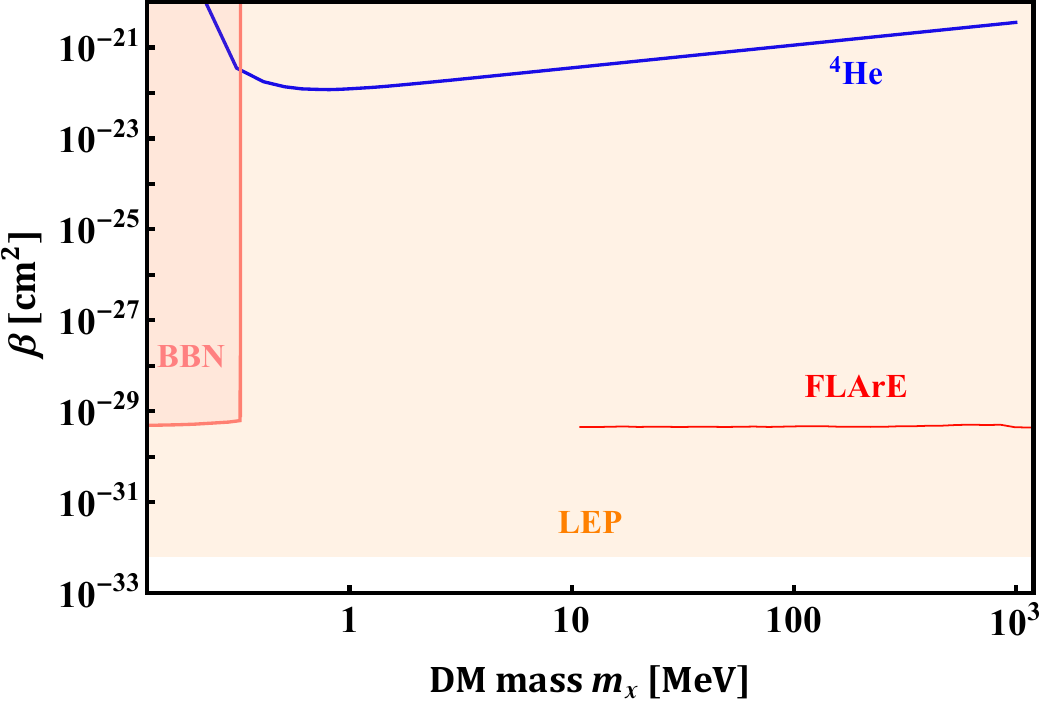}
		\end{minipage}%
		\begin{minipage}[c]{0.5\textwidth}
			\includegraphics[height=6cm,width=8cm]{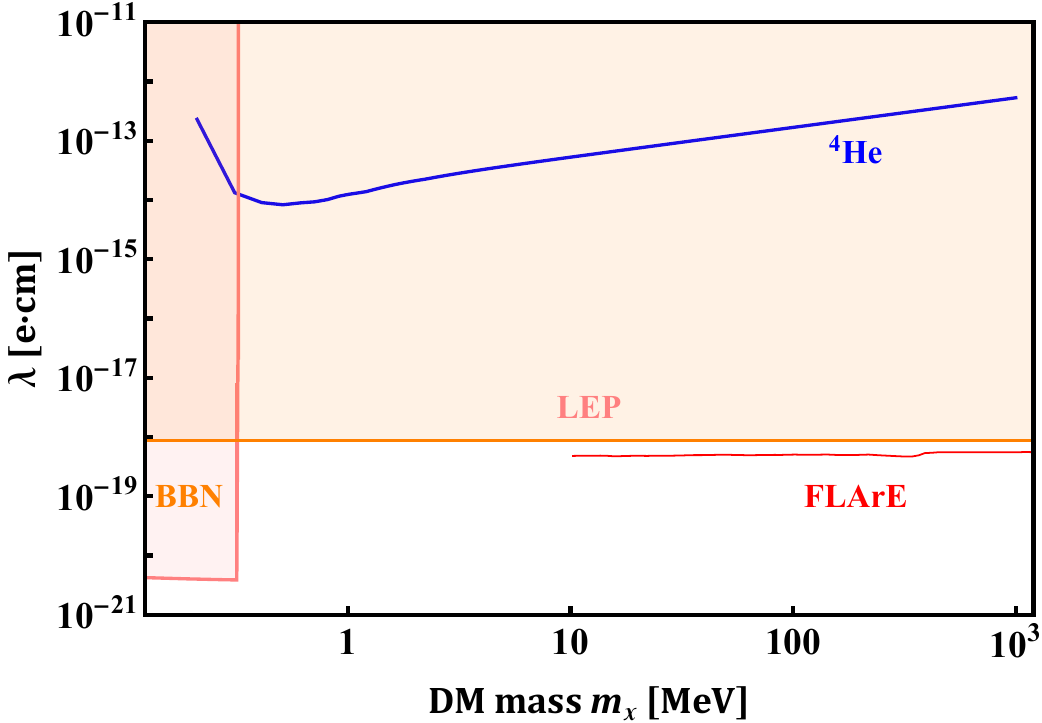}
		\end{minipage}
		\caption{Projected constraints on  anapole moment $\beta$ (left-panel) and magnetic dipole moment $\lambda$ (right-panel), respectively. We consider the 95$\%$ C.L. exclusion limit for  1 kg$\cdot$year exposure  of the superfluid $^4$He, assuming zero background. Shaded constraints are taken  from LEP~\cite{Chu:2018qrm}, FLArE~\cite{Kling:2022ykt} and BBN~\cite{Chu:2019rok}, respectively.} \label{fig:form factor}
	\end{center}
\end{figure}

By setting $E=100~{\rm kV/cm}$, we also show the  95$\%$ C.L. projected constraints on the anapole moment and the magnetic moment of the DM with $1$ kg$\cdot$year exposure in the left-panel and the right-panel of the Fig. ~\ref{fig:form factor}, respectively.
Although the exclusion limits on the charge radius are better than the limit on the anapole moment, the superfluid does not give a strong limit on these parameters, which is due to the limitation of superfluid material and detection technology. 
Today, the detection technology of the superfluid has difficulty reaching the meV order energy of a phonon, so for non-relativistic dark matter, the detection limit of a superfluid is theoretically on the keV scale DM.
When the mass of DM is less than a few hundred keV, due to the limitations of the detection threshold of phonon in superfluids, we still cannot give any limit on the  DM Electromagnetic form factors even if considering the use of quantum evaporation. With the improvement of detection technology, superfluid ${}^4$He can theoretically limit the parameters of DM with lower energy.

\section{ALP in the superfluid $^4$He}
	
In this section, we consider the phonon signal induced by the ALP in the superfluid. Instead of considering the effect of the  ALP-photon-photon interaction, whose coupling is constrained by various cavity experiments or direct detection experiments, we consider the effect of the axion-photon-dark photon interaction, 
\begin{equation}
	{\cal L} = {{{g_{a\gamma \gamma '}}} \over 2}a{F_{\mu \nu }}{\tilde F'^{\mu \nu }},\label{eq:L1}
\end{equation}
where  ${F_{\mu \nu }}$ and ${\tilde F'^{\mu \nu }}$ are the field strengths tensor for photon and dual field strengths tensor for dark photon. $a$ is the axion field. $g_{a\gamma \gamma '}$ is a  model-dependent coupling factor. The dual field tensor is given by ${\tilde F'^{\mu \nu }} = {1 \over 2}{\epsilon ^{\mu \nu \alpha \beta }}{F'_{\alpha \beta }}$,
where ${\epsilon ^{\mu \nu \alpha \beta }}$ is the Levi-Civita totally antisymmetric tensor. 

According to the Lagrangian in Eq.~\eqref{eq:L1}, we can find the Feynman rules for the conversion of photon-phonon by an external electric field- $\boldsymbol{E}$~\cite{Caputo:2019xum} which has been given in the Eq.(3),  and the axion-photon-dark photon vertex:
\begin{subequations}
\begin{align}
\includegraphics[valign=c,width=0.25\textwidth]{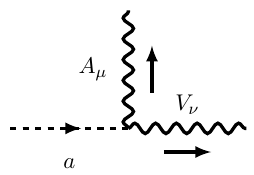} \!\!&=
{{{g_{a\gamma \gamma '}}} }{\epsilon ^{\mu \alpha \nu \beta }}{q_\mu }{{p'}_\nu }, \label{eq:rule2} 
\end{align}
\end{subequations}
where the crossed circle represents the external electric field. And $\omega_q$ and ${q_i}$ are the energy and momentum of the photon.  With the help of these Feynman rules, we can calculate the phonon signal induced by the ALP.

\subsection{single phonon emission with massive dark photon}
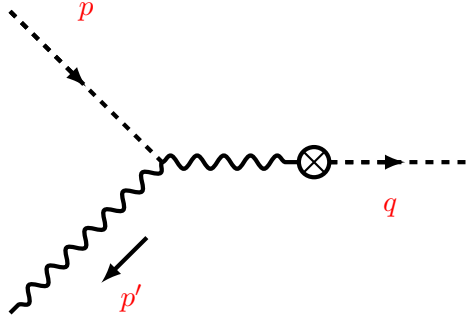
\begin{figure}
	\begin{center}
		\begin{tikzpicture}[scale=2]
			\draw[dashed,ultra thick] (0,2)--(1,1);
			\draw[-,snake=snake,ultra thick] (1,1)--(0,0);
			\draw[-latex,dashed,ultra thick] (0,2)--(0.5,1.5) ;
			\draw[-latex,ultra thick] (0.9,0.5)--(0.6,0.2) ;
			\draw[-,snake=snake, ultra thick] (1,1) -- (1.9,1);
			\draw[-,ultra thick] (2,1) circle (0.1) ;
			\draw[-,  thick, rotate around={45:(2,1)}] (1.9,1) -- (2.1,1);
			\draw[-,  thick, rotate around={-45:(2,1)}] (1.9,1) -- (2.1,1);
			\draw[-latex,dashed,ultra thick] (2.1,1) -- (2.6,1);
			\draw[dashed, ultra thick] (2.1,1) -- (3,1);
			\node[red, ultra thick] at (0.5,2) {$p$};
			\node[red, ultra thick] at (0.8,0.1) {$p'$};
			\node[red, ultra thick] at (2.5,0.7) {$q$};
		\end{tikzpicture}
		\caption{Feynman diagram for the amplitude of emission of a single phonon through the axion decay to photon and dark photon.} \label{fig:feynpro-arrprime}
	\end{center}
\end{figure}	
We first focus on the phonon signal in the case of massive dark photons. An ALP,  passing through the superfluid in the presence of an external electric field, scatters off $^4$He into a dark photon and a phonon. This process can be described by the Feynman diagram in Fig.~\ref{fig:feynpro-arrprime}.  A direct calculation gives
\begin{equation}
	i{\cal M} =  - i{{g_{a\gamma \gamma '}} \over {{q^2}}}{{\bar n{\alpha _E}} \over {{m_{He}}{c_s}}}\sqrt {{{{m_{He}}} \over {\bar n}}} \omega{\epsilon ^{\mu r\alpha s}} {q_\mu }{ p'_\alpha }\hat \varepsilon'^{*}_s{T_r}.
\end{equation}
Average the polarization of initial incoming particles and sum over their final polarization, ones gets 
\begin{equation}
	\begin{aligned}
		\sum\limits_{{\rm{spin}}} | M{|^2} = {{g_{a\gamma \gamma '}^2} \over {{q^4}}}{{\bar n\alpha _E^2} \over {{m_{He}}c_s^2}}{\omega ^2}\Bigg\{ {q^2}{T^2}{(p - q)^2} - {\Big[q \cdot (p - q)\Big]^2}{T^2} - {q^2}{\Big[(p - q) \cdot T\Big]^2}\Bigg\} ~.
	\end{aligned}
\end{equation}
The energy conservation shows that  $E'=E-\omega$, and the dispersion relation of phonon is $\omega  \simeq {c_s}|\textbf{q}|$. The angle between the incident axion and the outgoing phonon is taken as $\theta$. 
According to the analysis in the last section, one has
$dcos\theta =\frac{E-\omega}{|\textbf{k}||\textbf{q}|}d\omega$.
Integrating the final state phase space, we can express the differential rate as
%\begin{equation}
%	{{d\Gamma } \over {d\omega }} = ({{{{p'}^2}} \over {m_ \vee ^2}} - 1){{{g_{a\gamma \gamma '}^2}\bar n\alpha _E^2} \over {128\pi {m_a}{\omega ^2}{m_{He}}E{v_a}}}[{q^2}{(p - q)^2}{T^2} - {q^2}{(p \cdot T)^2} - {(q \cdot (p - q))^2}{T^2}]~.
%\end{equation}

%\begin{equation}
	%\begin{aligned}
	%{{d\Gamma } \over {d\omega }} =& {{g_{a\gamma \gamma '}^2\bar n\alpha _E^2} \over {16\pi {m_a}{\omega ^2}{m_{He}}E{v_a}}}[{q^2}{T^2}{(p - q)^2} - {(q \cdot p - q^2)^2}{T^2} - {q^2}{((p-q) \cdot T)^2}]
    %\end{aligned}
%\end{equation}
\begin{equation}
	\begin{aligned}
		{{d\Gamma } \over {d\omega }} =& {{g_{a\gamma \gamma '}^2\bar n\alpha _E^2} \over {16\pi {m_a}{\omega ^2}{m_{He}}E{v_a}}}{{|{\bf{E}}{|^2}{\omega ^2}} \over {c_s^2}}\Bigg\{ ({\cos ^2}{\theta _E} - c_s^2){\omega ^2}\left(1 - {1 \over {c_s^2}}\right)\Big[{E^2}(1 - v_a^2) 
		\\&+ 2E\omega\left( {{{v_a}} \over {{c_s}}}\cos \theta  - 1\right) + {\omega ^2}\left(1 - {1 \over {c_s^2}}\right)\Big] - {\omega ^2}\left(1 - {1 \over {c_s^2}}\right){(E\cos {\theta _E} - |{\bf{p}}{|^2}{c_s}\cos {\theta _a})^2} 
		\\&- ({\cos ^2}{\theta _E} - c_s^2){\Big[\omega \left(E - {{|{\bf{p}}|} \over {{c_s}}}\cos \theta \right) - {\omega ^2}\left(1 - {1 \over {c_s^2}}\right)\Big]^2}\Bigg\}  .\label{eq:differential rate2}
	\end{aligned}
\end{equation}
To obtain numerical results, we provide expressions of the four-momentums shown in the last equation, $q=(\omega,\textbf{q}), p=(E,\textbf{p}), T=(\textbf{E}\cdot\textbf{q},\omega\textbf{E}), p'=p-q=(E-\omega,\textbf{p}-\textbf{q})$.
Furthermore, we define the angles ${\theta _a}$ and ${\theta _E}$ as the angles between the incident axion and the electric field, and the emitted phonon and the electric field, respectively.

% The events rate per unit target mass is then 
%\begin{equation}
%	R=\int dv_\chi\,f_\text{MB}(v_\chi)\frac{\rho_\chi}{m_\text{He}\bar n m_\chi}\int_{\omega_\text{min}}^{\omega_\text{max}}d\omega \frac{d\Gamma}{d\omega}\,, \label{eq:R}
%\end{equation}
%Here we take $\rho_\chi=0.4\,~\text{GeV}/\text{cm}^3~$\cite{Trickle:2020oki}.
\begin{figure}[t]
	\centering
	\includegraphics[height=9cm,width=12cm]{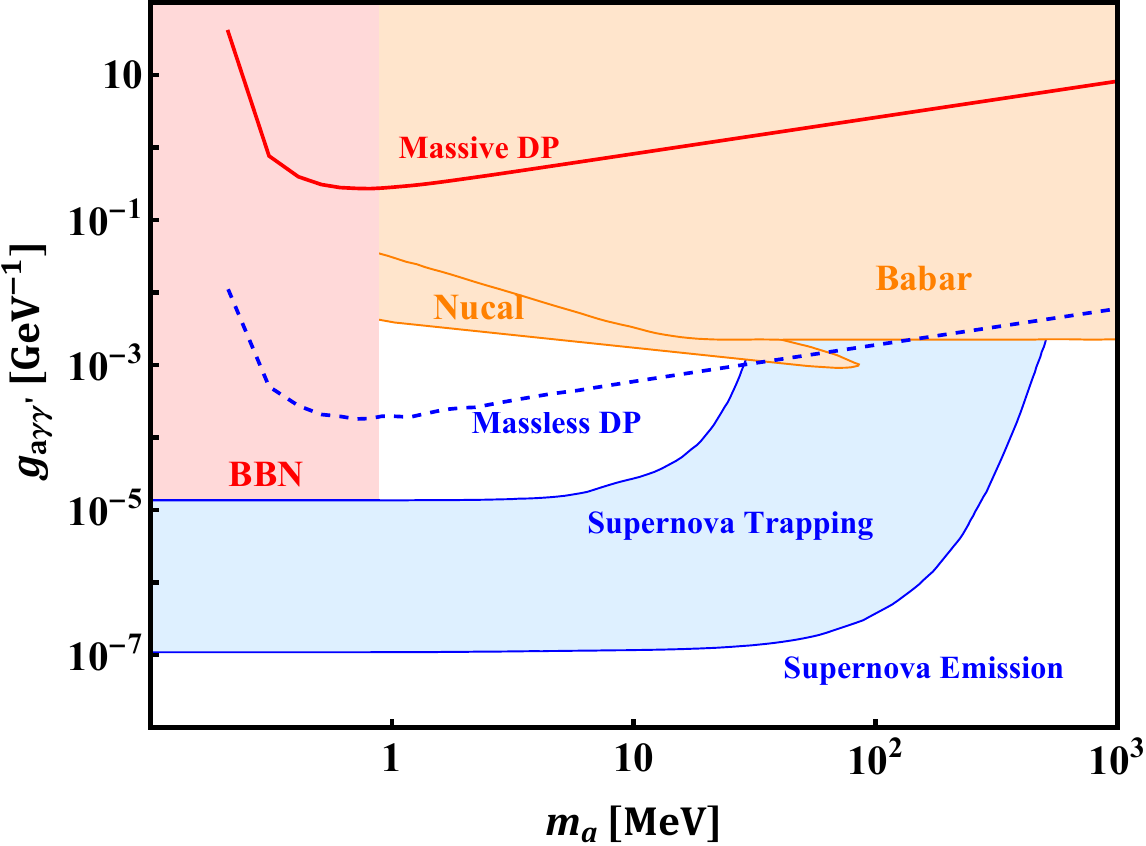}
	\caption{Projected constraints on  $g_{a\gamma \gamma '}$ with the massive dark photon (solid line) case and the massless dark photon (dashed line) case, respectively.  We consider the 95$\%$ C.L. exclusion limit for 1 kg$\cdot$year exposure of the superfluid, assuming zero background. Other shaded regions are constraints of BBN~\cite{deNiverville:2018hrc}, supernova~\cite{deNiverville:2018hrc}, Babar~\cite{BaBar:2001yhh,BaBar:2013agn} and Nucal~\cite{Dobrich:2019dxc,Blumlein:2013cua}, respectively. } 
	\label{fig:g-arr-arrprime}
\end{figure}
%\colorbox{red}{delete}The maximum energy transferred to a phonon by DM is determined either by kinematics (specifically by requiring $\text{cos}\theta  < 1$) or by the cutoff of the EFT, denoted as $\omega_\text{max}=\min\big(2E c_s(v_\chi-c_s),c_s\Lambda\big)$, ${c_s}\Lambda  \simeq 1~\text{meV}$.

%The energy emitted by a single phonon is insufficient for detection using calorimetric techniques, which typically have a sensitivity of (at best) 1 $\text{meV}$ \cite{Hochberg:2015pha}. Subsequently, it can lead to the evaporation of helium atoms, which can eventually be observed \cite{Hertel:2018aal,Maris:2017xvi}.  To achieve this, the phonon {\color{red}{also needs to}} overcome the binding energy between ${}^4$He atoms and the remaining bulk material, $\omega_\text{min}$ = 0.62 meV \cite{Maris:2017xvi}. For ${}^4$He atoms, the maximum energy of a phonon is approximately 1 meV, beyond this point, the dispersion relation is no longer linear. Specifically, given the value of $\omega_\text{max}$, this implies that only when the DM mass is approximately greater than 0.1 MeV, the final state phonons can be detected.

\subsection{single phonon emission with massless dark photon}

For a massless dark photon, the contraction of the scattering amplitude and its squared modulus are the same as these given in the last subsection, up to the replacement $p^{\prime 2} \to 0$,
\begin{equation}
	\begin{aligned}
		\sum\limits_{{\rm{spin}}} | M{|^2} = {{g_{a\gamma \gamma '}^2} \over {{q^4}}}{{\bar n\alpha _E^2} \over {{m_{He}}c_s^2}}{\omega ^2}\Bigg\{  - {q^2}{\Big[(p - q) \cdot T\Big]^2} - {(q \cdot p - {q^2})^2}{T^2}\Bigg\} ~.
	\end{aligned}
\end{equation}
We then find the differential emission rate to be 
\begin{equation}
	\begin{aligned}
		{{d\Gamma } \over {d\omega }} = &{\rm{ }}{{g_{a\gamma \gamma '}^2\bar n\alpha _E^2} \over {16\pi {m_a}{\omega ^2}{m_{He}}E{v_a}}}{{|{\bf{E}}{|^2}{\omega ^2}} \over {c_s^2}}\Bigg\{  - {\omega ^2}\Big(1 - {1 \over {c_s^2}}\Big){(E\cos {\theta _E} - |{\bf{p}}{|^2}{c_s}\cos {\theta _a})^2} \\
		&- ({\cos ^2}{\theta _E} - c_s^2){\Big[\omega \Big(E - {{|{\bf{p}}|} \over {{c_s}}}\cos \theta \Big) - {\omega ^2}\Big(1 - {1 \over {c_s^2}}\Big)\Big]^2}\Bigg\} .\label{eq:differential rate3}
	\end{aligned}
\end{equation}
It is worth mentioning that we also computed the process of an axion decaying to two photons and emitting a single photon. Its differential event rate aligns with the massless dark photon scenario, as given by Eq.~\eqref{eq:differential rate3}.

%By the rate in Eq.~\eqref{eq:differential rate2} and Eq.~\eqref{eq:differential rate3}, we can calculate projected constraints on the coupling factor $g_{a\gamma \gamma '}$ with massive scenario and massless scenario, respectively. Fig.~\ref{fig:g-arr-arrprime} shows the relationship between the coupling coefficient $g_{a\gamma \gamma '}$ and the axion mass. Specifically, we will examine an external electric field of $\textbf{E}$ = 100 kV/cm, a magnitude that has been demonstrated to be feasibly attainable in laboratory settings~\cite{Ito:2015hwa}. 
%lllustrated in Fig.~\ref{fig:g-arr-arrprime}, materials composed of superfluid ${}^4$He exhibit effective confinement capabilities for the coupling coefficient 
%$g_{a\gamma \gamma '}$ in the sub-MeV range. Among them, the massless scenario demonstrates superior limiting capabilities for the coupling coefficient $g_{a\gamma \gamma '}$ compared to the massive scenario, with a discrepancy of approximately three orders of magnitude.

Using the rates in Eq.~\eqref{eq:differential rate2} and Eq.~\eqref{eq:differential rate3}, we can calculate projected constraints on the coupling $g_{a\gamma \gamma '}$. 
Fig.~\ref{fig:g-arr-arrprime} shows the exclusion limit on the $g_{a\gamma\gamma^\prime}^{}$ as the function of the ALP mass at the 95\% C.L.  with 1 kg$\cdot$year exposure, where the solid and dashed lined correspond to the massive dark photon and the massless dark photon cases, respectively. 
For the massive dark photon case, we set its mass equal to that of the axion for simplicity. 
In particular, we set the external electric field strength to be $\textbf{E}$ = 100 kV/cm,
a magnitude that has been experimentally demonstrated to be achievable in laboratory settings~\cite{Ito:2015hwa}. As depicted in Fig.~\ref{fig:g-arr-arrprime}, superfluid ${}^4$He exhibits effective confinement capabilities for the coupling $g_{a\gamma \gamma '}$ within the sub-MeV mass range. Among these scenarios, the massless dark photon case exhibits enhanced constraining capabilities for the coupling  $g_{a\gamma \gamma '}$ in comparison to the massive dark photon scenario, with a difference of roughly three orders of magnitude. Other shaded regions are constraints of BBN~\cite{deNiverville:2018hrc}, supernova~\cite{deNiverville:2018hrc}, Babar~\cite{BaBar:2001yhh,BaBar:2013agn} and Nucal~\cite{Dobrich:2019dxc,Blumlein:2013cua}, respectively.

\section{Discussion}

Condensed matter materials have shown great potential in detecting sub-GeV DM. 
In this paper, we have studied the detectability of the superfluid $^4$He on two kinds of sub-GeV DM models: Fermionic DM with non-zero electromagnetic form factors and the ALP that couples to the photon and a dark photon.  Using the EFT of the superfluid, we calculated the event rate of phonon induced by the charge radius, the anapole, and the magnetic moment of the Fermionic DM, respectively, and the analytical results are given in both relativistic and non-relativistic limits.  Numerical results show that competitive constraints on these couplings can be derived with 1 kg$\cdot$year exposure in a projected experiment, which provides an important reference for other experimental detections.  For the ALP-photon-dark-photon coupling, which cannot be detected in  Helioscope or Haloscope, we have derived the analytical results for the ALP scattering into a dark photon and a phonon. Then we have shown constraints on this coupling in a projected experiment, which is the first direct detection constraint on this coupling. Since the nuclei of helium are relatively light, we can also measure high-momentum, low-mass axion flow. One can also use the superfluid cavity~\cite{Baker:2023kwz}  to probe axion, this is our next research work. 

It should be mentioned that there is still a lot of room for the choice of materials. Materials such as graphene~ \cite{Kim:2020bwm} may offer lower detection limits. We can use multiple excitations to detect more targets in different cases~\cite{Knapen:2016cue}. Target Anisotropies and Daily Modulation~\cite{Trickle:2019nya} 
have not been considered in limiting DM Electromagnetic form factor. The results will be more realistic with those considerations. We may take this as a future work.
One can also study DM-superfluid helium interaction by condensed matter approach with dynamic structure factor~\cite{Kahn:2021ttr,Baym:2020uos}.
The use of the cascade process of a dark matter particle with a mass between 1 MeV and 1 GeV scattering off superfluid ${}^4$He has already been considered~\cite{You:2022pyn}. In contrast, our method is a theoretical study in the context of particle physics, which offers a direct and concise picture of the processes ~\cite{Campbell-Deem:2022fqm,Mitridate:2021ctr,Coskuner:2021qxo}.

\vspace{1cm}
	\begin{acknowledgments}
	This work was supported by the National Natural Science Foundation of China under grant No. 11775025,12105013 and the Fundamental Research Funds for the Central Universities under grant No. 2017NT17.
\end{acknowledgments}

%%%%%%%%%%%%%%%%%%%%%%%%%%%%%%%%%%%%%%%%%%%%%%
%%%%%%%%%%%%%%%%%%%%%%%%%%%%%%%%%%%%%%%%%%%%%%
%%%%%%%%%%%%%%%%%%%%%%%%%%%%%%%%%%%%%
\bibliography{references}

\end{document}